\documentclass[prl,aps,floatfix,showpacs,twocolumn]{revtex4}
\usepackage{dcolumn}
\usepackage{bm}
\usepackage{color}
\usepackage{graphicx}
\usepackage{epsf}
\usepackage{pstricks}
\def\neff{N_{\rm eff}}
\newcommand{\be}{\begin{equation}}
\newcommand{\ee}{\end{equation}}
\newcommand{\bea}{\begin{eqnarray}}
\newcommand{\eea}{\end{eqnarray}}

\begin{document}
\title{Implication of the proton-deuteron radiative capture for  
Big Bang Nucleosynthesis}
\author{L.E.\ Marcucci$^{\, {\rm a,b}}$, 
G.\ Mangano$^{\,{\rm c}}$,
A.\ Kievsky$^{\,{\rm b}}$, and M.\ Viviani$^{\,{\rm b}}$}
\affiliation{
$^{\,{\rm a}}$\mbox{Department of Physics, University of Pisa, Largo B. Pontecorvo 3, I-56127 Pisa, Italy}\\
$^{\,{\rm b}}$\mbox{INFN-Pisa, Largo B. Pontecorvo 3, I-56127 Pisa, I-56127 Pisa, Italy}\\
$^{\,{\rm c}}$\mbox{INFN-Napoli, Complesso Univ. Monte S. Angelo, 
Via Cintia, I-80126 Napoli, Italy}\\
}

\date{\today}

\begin{abstract}
The astrophysical $S$-factor for the radiative capture $d(p,\gamma)^3$He 
in the energy-range of interest for Big Bang Nucleosynthesis (BBN)
is calculated using an {\it ab-initio} approach. The 
nuclear Hamiltonian retains both two- and three-nucleon
interactions - the Argonne $v_{18}$ and the Urbana IX, respectively.
Both one- and many-body contributions
% (the so-called
%meson-exchange currents) 
to the nuclear
current operator are included. The former retain for the first time, 
besides the $1/m$ leading order contribution ($m$ is the nucleon mass), 
also the next-to-leading order term, proportional to $1/m^3$. The many-body
currents are constructed in order to satisfy the current conservation
relation with the adopted Hamiltonian model. The hyperspherical
harmonics technique is applied
to solve the $A=3$ bound and scattering states.
A particular attention is used in this second case in order to obtain, in the
energy range of BBN, an uncertainty on the astrophysical $S$-factor
of the order or below $\sim$1 \%. Then, in this energy range,  
the $S$-factor is found to be 
$\sim$10 \% larger than the currently adopted values.
Part of this increase (1--3 \%) 
is due to the $1/m^3$ one-body operator, while the remaining is due to 
the new more accurate scattering wave functions.
We have studied the implication of this new determination 
for the $d(p,\gamma)^3$He $S$-factor on
deuterium primordial abundance. We find that the predicted theoretical value 
for $^2$H/H is in excellent agreement with its experimental determination, 
using the most recent determination of baryon density of Planck experiment, 
and with a standard 
number of relativistic degrees of freedom $\neff=3.046$ during primordial 
nucleosynthesis.

\end{abstract}


\index{}\maketitle

{\it Introduction.}
The radiative capture $d(p,\gamma)^3$He is a relevant process in many
astrophysical environments. For instance, 
it is the second step in the chain of nuclear reactions which, 
starting from the proton-proton weak capture, allows to stars like our Sun
to shine via $pp$-chain. Interest in this reaction 
is also present  in the context of Big Bang Nucleosynthesis (BBN),  
see e.g. Ref.~\cite{Ioc09} and references therein, since it is one of the main processes
through which deuterium can be destroyed and thus affects its eventual yield. BBN is a powerful method to test the validity of the cosmological
model at the MeV energy scale. For a given set of values of leading nuclear reaction rates, primordial element
abundances depend upon two key cosmological parameters, 
the energy density in baryons, $\Omega_bh^2$, and the energy density of relativistic species $\rho_{rel}$ or equivalently, the
effective neutrino number, $\neff$, defined as
%\begin{equation}
$\rho_\mathrm{rel}=  \rho_\gamma \left(1+ \frac{7}{8}\, \left( \frac{4}{11} \right)^{4/3} \neff \right)$ 
%\ , \label{eq:neffdef} 
%\end{equation} 
where $\rho_\gamma$ is the Cosmic Microwave Background (CMB) photon energy density. The benchmark value for this parameter with only three active neutrinos contributing is $\neff=3.046$ \cite{Mangano:2005cc}.

The baryon energy density has been measured with a remarkable precision by CMB anisotropy experiments, so in the minimal $\Lambda$CDM scenario light nuclei abundances are basically fixed~\cite{Planck2015}. 
Using the most recent determination of the Planck satellite experiment, 
$\Omega_bh^2=0.02225 \pm 0.00016$~\cite{Planck2015}, and 
the public BBN code \texttt{PArthENoPE}~\cite{Pis08}, 
the primordial deuterium to hydrogen ratio is predicted to be
$^2$H/H=$(2.60 \pm 0.07)\times 10^{-5}$ (68 \% C.L. error). Notice that this determination is lower than before, due to the order 1\% increase of $\Omega_b h^2$ with respect to Planck 2013 results. The theoretical uncertainty quoted above is almost fully due (more than 90 \%) to present errors on experimental determination of the $d(p,\gamma)^3$He cross section.

This result is in agreement at 1-$\sigma$ with the recent determination of Ref.~\cite{Coo14} where, 
through a new analysis of all known deuterium absorption-line systems, it 
was found $^2$H/H=$(2.53 \pm 0.04)\times 10^{-5}$ at 68 \% of C.L.,
but it is slightly larger. A possible way to get an even better agreement 
between the two values is to slightly decrease the effective neutrino number down to
$\neff \sim 2.84$.
Another way is to increase the value of the $d(p,\gamma)^3$He 
astrophysical $S$-factor. This possibility was first explored in 
Ref.~\cite{Nollett:2011aa} and then analyzed in details in  
Ref.~\cite{DiValentino:2014cta}, using Planck 2013 data release, and in
Ref.~ \cite{Planck2015}. The conclusion of these studies is that increasing 
the  $d(p,\gamma)^3$He thermal rate in the BBN temperature range by a factor 
of order 10 \% leads to a very good agreement between CMB anisotropy results 
and primordial deuterium abundance.
Therefore, a better determination of this $S$-factor 
with a reduction of the corresponding uncertainty 
in the BBN energy range, i.e. $E\simeq 30-300$ keV,  
would be extremely important.
% to asses the overall compatibility of a standard BBN scenario 
%in a minimal $\Lambda$CDM model with the result of Ref.~\cite{Coo14}. 

The astrophysical $S$-factor at low
energy, around the solar Gamow peak $E_{G}\simeq 9$ keV, is 
well known, thanks to the results of the LUNA experiment~\cite{LUNA02}.
However, for the BBN relevant energy range, the experimental situation
is rather unclear, since the only available experimental data~\cite{Ma97} 
are quite in disagreement with the polynomial best fit of $S(E)$
for $E\simeq 0-2$ MeV~\cite{Ade11}. 
This gives rise to an uncertainty on the cross 
section at the level of 6--10 \%. This is the main motivation 
behind the experiment recently proposed by the LUNA Collaboration, which
has the goal of measuring the $d(p,\gamma)^3$He astrophysical $S$-factor in the
BBN energy range with a 3 \% accuracy. A feasibility test has already been 
performed~\cite{TrezziPV}.

On the other hand, the $d(p,\gamma)^3$He astrophysical $S$-factor 
can be calculated using a microscopic {\it ab-initio} 
approach. In fact, in Refs.~\cite{Viv00,Mar05} (see also references therein)
the hyperspherical harmonics (HH) technique was used 
to solve for the $A=3$ nuclear wave functions using a realistic description 
of the nuclear interaction, which includes both two- and three-nucleon 
interactions. These are constructed to reproduce the $A=2$ large body 
of experimental data with a $\chi^2$/datum $\sim$ 1 (the Argonne
$v_{18}$ - AV18 - model~\cite{Wir95}), 
and the $A=3$ binding energies (the Urbana IX - UIX -
model~\cite{Pud97}). Note
that the HH method (see Ref.~\cite{Kie08} for a review) 
is the only available one able to calculate
the nuclear wave function for the initial $p-d$ scattering state at low 
energies, as the ones of interest here, including the Coulomb interaction
between the charged initial particles. In the latest study of 
Ref.~\cite{Mar05}, a realistic
model for the nuclear current operator was used, retaining both one-
and many-body contributions. The latter are necessary in order
to maintain gauge invariance in the presence of a system of interacting
particles, and in Ref.~\cite{Mar05} all the effort was put to construct
these contributions which exactly satisfy the current conservation
relation in conjunction with the AV18/UIX potential. The former, instead, 
was simply obtained performing a $1/m$
expansion ($m$ is the nucleon mass) of the single-nucleon covariant
operator, and retaining the leading order contribution. 
The many-body contributions
were found already in Ref.~\cite{Viv00} 
essential to reach an excellent agreement with
the LUNA data~\cite{LUNA02} around the solar Gamow peak~\cite{Mar05,Ade11}. 
In the energy range of interest for BBN, on
the other hand, the theoretical predictions
of Ref.~\cite{Mar05} were found to be 2--10 \%
higher than the central value for the 
polynomial fit of Ref.~\cite{Ade11}. 
In the present letter, our starting point is the work of Ref.~\cite{Mar05},
which, although very accurate, should be considered incomplete for two 
reasons: (i) no estimate of the theoretical uncertainty 
was given, in particular that one arising from the solution of the $p-d$ 
scattering problem with the HH method; (ii) the one-body terms
beyond the leading order operator, of the order $1/m^3$, 
were found few years later~\cite{Gir10} essential
in order to get a reasonable agreement between theory and experiment
for a related process, the $d(n,\gamma)^3$H radiative capture. Given the
similarities between the $p-d$ and $n-d$ radiative captures, it is to 
be expected that these $1/m^3$ one-body contributions might be
important also for the process here under consideration. 
The goal of the present letter is to address the two
above mentioned issues and to verify whether the new prediction
for the $d(p,\gamma)^3$He astrophysical $S$-factor goes in the direction of improving the consistency of theoretical BBN deuterium abundance prediction,  the new Planck results, and the experimental data of Ref.~\cite{Coo14}. We do not consider here $^4$He primordial mass fraction $Y_p$, since it is insensitive to
this reaction rate.  For example, a change of the $d(p,\gamma)^3$He $S$-factor by a factor two affects $Y_p$ at the level of 0.04 \%, too small to be appreciated with present statistical and systematic uncertainties on its experimental determination, see e.g. Ref.~\cite{Cyburt:2015mya}.

{\it The present calculation.} We discuss here the two 
significant improvements
in the calculation with respect to Ref.~\cite{Mar05}. First of all,
in the present work, we pay particular attention to the numerical
determination of the $p-d$ wave function. In particular, in each 
$L,S,J$ channel ($L$ is the $p-d$ orbital angular momentum, 
$S=1/2,3/2$ is the $p-d$ total spin, and
${\bf J}={\bf L}+{\bf S}$)
the wave function $\Psi^{LSJ}$  has been tested calculating
$\langle H \rangle \equiv\langle \Psi^{LSJ} | H | \Psi^{LSJ} \rangle$ 
in a box with a radius of 70 fm, 
using a Monte Carlo method (independently
on our technique to determine $\Psi$)
and verifying that the correct result $\langle H\rangle=E-B_d$ is obtained
within the requested accuracy (here $E$ is the 
$p-d$ center-of-mass energy and $B_d$ the deuteron binding energy). 
Different grid points,
dimensionality of the HH expansion basis and values for
the non-linear parameter 
entering the polynomial expansion of the hyperradial
functions (see Ref.~\cite{Kie08}) have been 
checked in order to verify the above relation within 0.1 \%. 
With this procedure we were able to reduce the numerical uncertainty relative
to the wave functions in our astrophysical $S$-factor estimates
of better
than 1 \% for the whole energy range here under consideration (see
Table~\ref{tab:res}). 
To be noticed, that 
the nuclear Hamiltonian used in the present study is the same as that
of Ref.~\cite{Mar05}, i.e., it retains the AV18/UIX potential model,
which allows to nicely reproduce, using the HH method, the
$^3$He binding energy and many $A=3$ scattering
observables. Some discrepancies between the AV18/UIX predictions and 
experiment arise only for some delicate polarization and breakup 
observables. 
Therefore, the shortcomings of the adopted interaction model are expected 
to be of no consequences for the present study.
Furthermore, 
all partial waves with $J\leq 5/2$ and both parities have been retained. 

The new term of the nuclear electromagnetic current here included
is that arising from the $1/m$ expansion of the single-nucleon
covariant current operator, and is a relativistic correction
of the order $1/m^3$. It has been
derived in Refs.~\cite{Pas09,Son09} in the context of
chiral effective field theory, and can be written as~\cite{Pas09}
%
%\begin{widetext}
\begin{eqnarray}
{\bf j}_i^{RC}&=&-\frac{e}{8m^3}{e_i}\biggl[
2\biggl(K_i^2+q^2/4\biggr)(2{\bf K}_i+{\rm i}{\bm\sigma}_i\times
{\bf q})
\nonumber \\
&+&
{\bf K}_i\cdot{\bf q}({\bf q}+2{\rm i}{\bm\sigma}_i\times{\bf K}_i)\biggr]
\nonumber \\
&-&
\frac{{\rm i}e}{8m^3}{\kappa_i}\biggl[{\bf K}_i\cdot{\bf q}
(4{\bm\sigma}_i\times{\bf K}_i-{\rm i}{\bf q})
\nonumber \\
&-&
(2{\rm i}{\bf K_i}-{\bm\sigma}_i\times{\bf q})q^2/2+
2({\bf K_i}\times{\bf q}){\bm\sigma}_i\times{\bf K}_i\biggr] ,
\label{eq:1brc}
\end{eqnarray}
%\end{widetext}
%
where ${\bf K}_i=({\bf p}_i^\prime+{\bf p}_i)/2$, ${\bf p}_i$ and
${\bf p}_i^\prime$ being the initial and final momenta of the nucleon,
${\bf q}$ is the photon momentum, $e$ is the electron charge,
${e_i}= (1+\tau_{i,z})/2$, the 
charge-projection isospin operator,
$\kappa_i=(\kappa_S-\kappa_V\tau_{i,z})/2$,
$\kappa_S=-0.12\mu_N$ ($\kappa_V=3.706\mu_N$) being the isoscalar
(isovector) combination of the anomalous magnetic moments of
proton and neutron, 
and ${\bm\sigma}_i$ (${\bm\tau}_i$) 
are the spin (isospin) Pauli matrices. It was found in Ref.~\cite{Gir10}
that ${\bf j}_i^{RC}$ reduces the $n-d$ total cross section at thermal energies
of about 4-5 \%, bringing the theoretical prediction in a much better agreement
with the experimental datum (within 4 \%). In the $p-d$ case, instead,
we have found that 
the operator ${\bf j}_i^{RC}$ gives a positive contribution, increasing
the astrophysical $S$-factor of 1--3 \% over the whole energy range considered
here (see Table~\ref{tab:res}).

\begin{table}[hbt]
\vspace*{-0.3cm}
\caption{\label{tab:res} The $p-d$ astrophysical $S$-factor (in keV b) for a 
representative set of energy values $E$ (in keV). The theoretical percent
uncertainty arising from the solution of the $p-d$ scattering problem 
with the HH method is given in the second column ($\Delta S_{WF}$), 
while the additional contribution due to the one-body term of 
Eq.~\protect(\ref{eq:1brc}) ($\Delta{\bf j}^{(RC)}$) is given in the last
column, also in percent. Note that for $E$=2 MeV (last row), the value for
$\Delta S_{WF}$ is below the permil level, and therefore not quoted.}
\begin{tabular}{c|c|cc}
$E$ [keV] & $S(E)$ [keV b] & $\Delta S_{WF}$ [\%] & $\Delta{\bf j}^{(RC)}$ [\%] \\
\hline
10 & 0.286 & 0.1 & +0.8 \\
20 & 0.355 & 1.0 & +1.1 \\
35 & 0.460 & 1.1 & +1.3 \\
50 & 0.570 & 0.9 & +1.7 \\
70 & 0.716 & 0.4 & +2.1 \\
95 & 0.912 & 0.3 & +2.3 \\
120 & 1.112 & 0.8 & +2.4 \\
145 & 1.317 & 0.4 & +2.5 \\
170 & 1.529 & 0.4 & +2.6 \\
195 & 1.748 & 0.4 & +2.6 \\
220 & 1.968 & 0.5 & +2.8 \\
245 & 2.197 & 0.4 & +2.7 \\
260 & 2.343 & 0.9 & +2.8 \\
300 & 2.716 & 0.5 & +2.7 \\
400 & 3.676 & 0.6 & +2.7 \\
500 & 4.739 & 0.2 & +2.7 \\
750 & 7.539 & 0.3 & +2.6 \\
1000 & 10.685 & 0.4 & +2.7 \\
2000 &  25.908 & -- & +2.3 \\
\end{tabular}
\end{table}
%\vspace*{0.8cm}
\begin{figure}[bth]
\includegraphics[height=10cm,width=8cm]{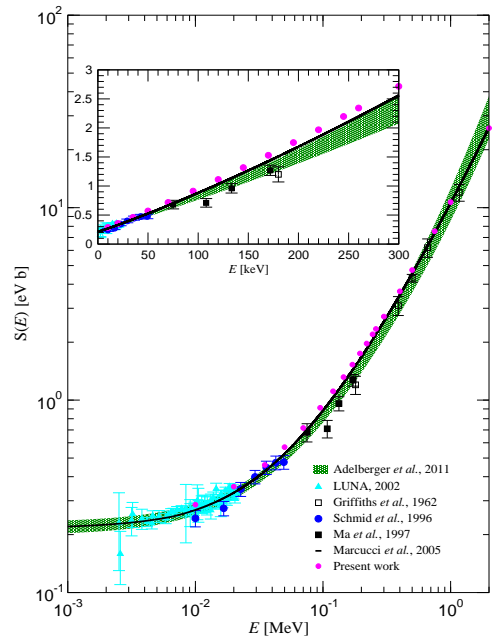}
%includegraphics[height=12cm,width=8cm]{Spd.v3.ps}
%\vspace*{0.5cm}
\caption{(Color online) 
The astrophysical $S$-factor obtained in the present work 
(magenta up-triangles) is plotted
together with the available experimental data of 
Refs.~\protect\cite{Gri62,Sch96,Ma97,LUNA02}, the calculation of 
Ref.~\protect\cite{Mar05} (solid black line), and the quadratic best
fit to the data of Ref.~\protect\cite{Ade11} (green band). The
inset shows the astrophysical $S$-factor in the 0-300 keV energy range,
of relevance for BBN.}
\label{fig:sfactor}
\end{figure}
%\vspace*{0.8cm}

The astrophysical $S$-factor obtained in the present work is 
listed in Table~\ref{tab:res} and plotted
in Fig.~\ref{fig:sfactor}, where it is compared with the previous
calculations of Ref.~\cite{Mar05}, as well as with the existing data 
of Refs.~\cite{Gri62,Sch96,Ma97,LUNA02} and
the polynomial best fit of Ref.~\cite{Ade11}. To be noticed that
the theoretical uncertainty arising from the solution of the $p-d$ 
scattering problem with the HH method are not visible on the plot, although
the corresponding symbols retain an error. The present results are 
systematically larger (about 8-10 \%) than those of Ref.~\cite{Mar05}. 
We have investigated
the origin of such an increase, and we have found that only 1-3 \%, depending 
on the energy value, is due to the one-body $1/m^3$ contribution. Therefore, 
the remaining 5-8 \% is due to the new solutions of the $A=3$ (scattering)
problem. In fact, the present wave functions have been obtained with the same 
HH technique as in Ref.~\cite{Mar05}, but with the goal 
of reaching a required higher accuracy, as dictated
by BBN, and therefore they have
been tested one by one, as explained above.

{\it Implications for BBN.} To study the effect of the new {\it ab-initio} 
determination of the $d(p,\gamma)^3$He $S$-factor on primordial deuterium 
produced during BBN we have computed  the corresponding thermal rate using 
the best fit values reported in the second column of Table \ref{tab:res} and 
modified the numerical code \texttt{PArthENoPE}~\cite{Pis08} accordingly. 
The theoretical results for deuterium to hydrogen density ratio  
$^2$H/H$_{th}$ are then computed as function of two parameters, the baryon 
density $\Omega_b h^2$ and $\neff$ and compared with the experimental 
determination $^2$H/H$_{exp}$ of Ref.~\cite{Coo14}. 
To obtain the best fit values 
and uncertainty on these parameters we then consider the likelihood function
\bea 
&& {\cal L}(\Omega_b h^2,\neff) =  \nonumber {\cal L}_{Planck}(\Omega_b h^2) 
\times \\ 
&& \exp \left(- \frac{(^2\mbox{H}/\mbox{ H}_{th}(\Omega_b h^2,\neff) -
\mbox{$^2$H}/\mbox{H}_{exp})^2}{2(\sigma^2_{exp} +\sigma^2_{th})} \right) 
\ ,  \label{likely}
\eea
where $\sigma_{exp}=0.04$ as obtained in Ref.~\cite{Coo14}. 
${\cal L}_{Planck}(\Omega_b h^2)$ is a Gaussian prior corresponding to the 
Planck result of Ref.~\cite{Planck2015}, 
while $\sigma_{th}$ is the propagated error on deuterium yield due to the 
present experimental uncertainty on other leading nuclear reactions relevant 
for deuterium production and destruction during BBN, namely  $d(d,n)^3$He and $d(d,p)^3$H. It also accounts for the 
$d(p,\gamma)^3$He $S$-factor theoretical uncertainty of Table~\ref{tab:res}, and the small error on the $p(n,\gamma)d$ rate,
both very subdominant with respect to the other error sources. 

%\vspace*{0.8cm}
\begin{figure}[bth]
\includegraphics[height=7.cm,width=7.cm]{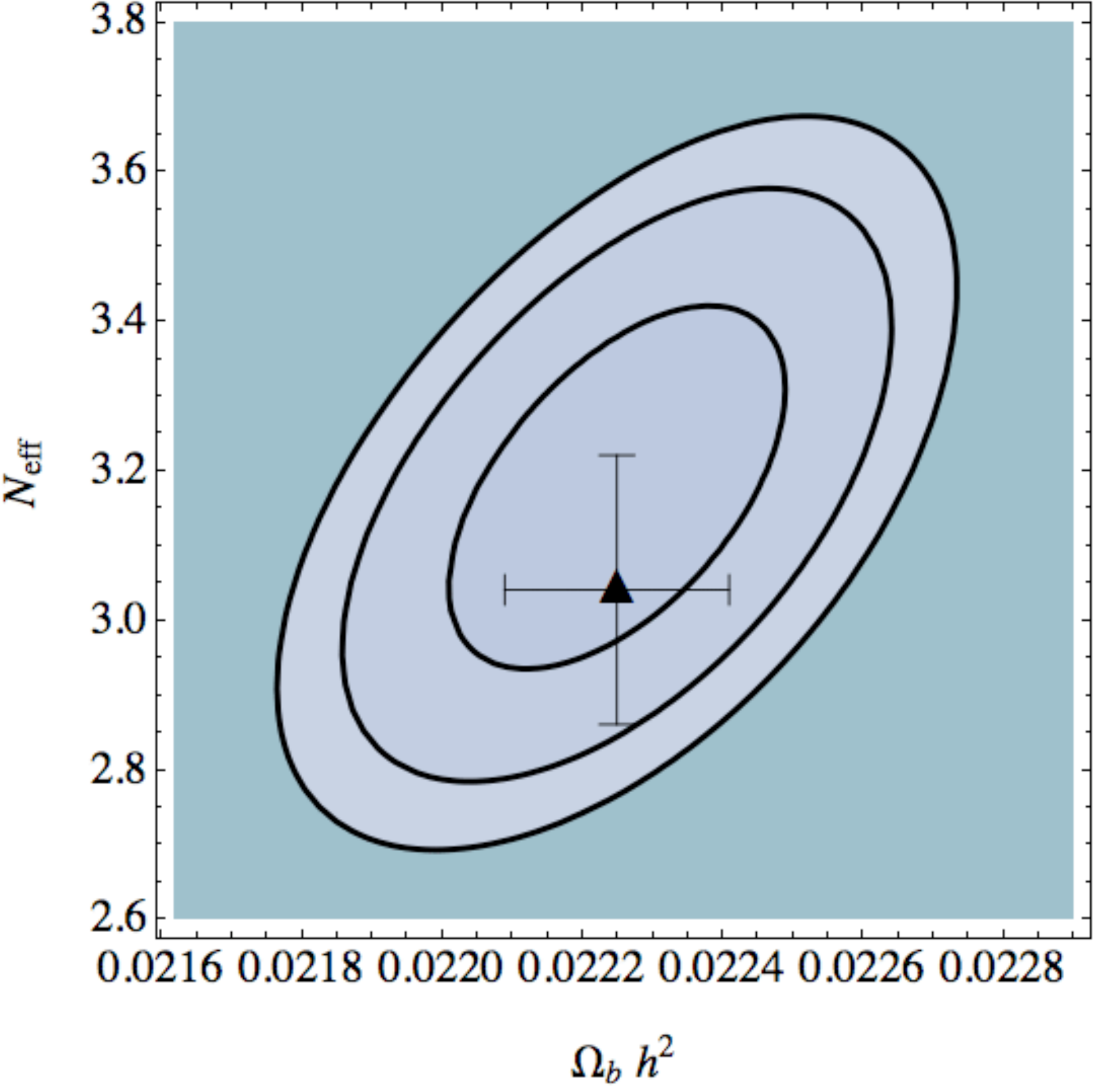}
%includegraphics[height=7cm,width=9cm]{dph3he.eps}
%\vspace*{0.5cm}
\caption{(Color online) 
The likelihood contours (68, 95 and 99 \% C.L.) in the $\Omega_b h^2$-$\neff$ 
plane from $^2$H/H, with the Planck 2015 prior on $\Omega_b h^2$, a free $\neff$ and using the experimental 
result of Ref.~\cite{Coo14}. The triangle is the best fit value of Planck 2015 
results for these parameters, with corresponding 68\% C.L. error 
bars~\cite{Planck2015}.}
\label{fig:likely}
\end{figure}
%\vspace*{0.8cm}
For the Planck 2015 value of $\Omega_b h^2$ and standard $\neff$ we get $^2\mbox{H}/\mbox{ H}_{th}=2.49 \pm 0.03 \pm 0.03$, where the two errors are due to nuclear rate and $\Omega_b h^2$ uncertainties, respectively.
In Fig.~\ref{fig:likely} we show the likelihood contours 
(68, 95 and 99 \% C.L.) in the $\Omega_b h^2 -\neff$ plane. 
The Planck 2015 best 
fit results and 68 \% C.L. error bars are also shown. The agreement is 
very good, within 1-$\sigma$. Marginalizing over baryon energy density, we 
find $\neff = 3.18 \pm 0.16\, (0.32)$, where the error is at 68 \% (95 \%) C.L., 
which is fully consistent with a standard radiation content during BBN. 
Notice that, once the Planck 2015 prior on $\Omega_b h^2$ is used, the 
uncertainty one gets on $\neff$ from  $^2$H/H alone is of the same order of 
magnitude obtained from CMB, once Baryon Acoustic Oscillation data are also 
exploited. In this case Planck result is $\neff= 3.04 \pm 0.18$ 
(68 \% C.L.)~\cite{Planck2015}. 

{\it Summary and Outlook.} The astrophysical $S$-factor for the
$d(p,\gamma)^3$He reaction is crucial to determine the consistency of  BBN theoretical prediction for deuterium abundance, the new Planck results, and the
most recent experimental determination of such abundance~\cite{Coo14}.
In the absence of an accurate experimental determination in the energy
range of interest for BBN, 30-300 keV, we have performed a new 
theoretical {\it ab-initio} calculation, using the most up-to-date
techniques to calculate the wave functions for the 
initial scattering and final bound states, with the realistic 
AV18/UIX potential model, 
as well as the most up-to-date realistic model for the 
nuclear current operator, which satisfies gauge invariance 
with the adopted Hamiltonian and retains
the $1/m^3$ contribution in the one-body operator. We have found that the
numerical uncertainty relative to the wave function in the $S$-factor
is lower than 1 \%, while the $1/m^3$ one-body contribution increases
the $S$-factor by 1--3 \% over the whole energy range. We have then
investigated the effect of this new
{\it ab-initio} determination on the primordial deuterium abundance. 
We find that BBN predictions are in very good agreement with the Planck
2015 results and the experimental result of ~\cite{Coo14}. Also the inferred value for
$\neff$ is fully consistent with a standard radiation content during BBN.
Of course, our results ought to be confirmed by direct measurement
of the $d(p,\gamma)^3$He $S$-factor, as it is planned at the 
Gran Sasso National Laboratories (Italy), by the LUNA Collaboration.
Such a measurement will therefore turn out to be crucial in this context and an improved accuracy at few percent would provide an independent handle to assess 
the overall consistency of the standard cosmological model.
Finally, it should be mentioned that a theoretical {\it ab-initio} calculation
of this  process is at reach also within the so-called
chiral effective field theory ($\chi$EFT) framework, 
which provides, on one hand, a direct connection between 
quantum chromo-dynamics (QCD) and the strong and electroweak interactions 
in nuclei. On the other hand, it is a practical calculational scheme, which 
can be improved systematically. This allows,
in principle, to obtain predictions on a more fundamental basis.
The first studies along this line have been done for 
the electromagnetic structure of light nuclei~\cite{Pia13,Mar15}.
At present, however, the consistency between the $\chi$EFT
nuclear potentials and electromagnetic currents necessary to satisfy 
{\it exactly} gauge invariance (as in the case of the calculation
presented here) has not
been yet achieved, making the $\chi$EFT results not completely
reliable at the accuracy level necessary for BBN predictions.
Work on this issue, though, is currently underway.

%\section*{Acknowledgments}
{\it Acknowledgments.} 
The authors acknowledge the support of the computer center staff at INFN-Pisa, 
where part of the present calculations were performed.
G.M. acknowledges support by
INFN
%{\it Instituto Nazionale di Fisica Nucleare} 
I.S. TASP and  PRIN 2010 
``Fisica Astroparticellare: Neutrini ed Universo Primordiale'' of the
Italian {\it Ministero dell'Istruzione, Universit\`a e Ricerca}. L.E.M.\
would like to thank the LUNA Collaboration for triggering the
present study and for useful discussions.


\begin{thebibliography}{100}
%
\bibitem{Ioc09}
F.\ Iocco, G.\ Mangano, G.\ Miele, O.\ Pisanti, and P.D.\ Serpico,
Phys.\ Rep.\ {\bf 472}, 1 (2009)
%
%\cite{Mangano:2005cc}
\bibitem{Mangano:2005cc} 
  G.~Mangano, G.~Miele, S.~Pastor, T.~Pinto, O.~Pisanti and P.~D.~Serpico,
  %``Relic neutrino decoupling including flavor oscillations,''
  Nucl.\ Phys.\ B {\bf 729}, 221 (2005)
  %%CITATION = HEP-PH/0506164;%%
  %
  \bibitem{Planck2015}
P.A.R.\ Ade {\it et al.} (Planck Collaboration), arXiv:1502.01589 [astro-ph.CO]
%
\bibitem{Pis08}
O.\ Pisanti, A.\ Cirillo, S.\ Esposito, F.\ Iocco, G.\ Mangano,
G.\ Miele, and P.D.\ Serpico, Comput.\ Phys.\ Comm.\ {\bf 178}, 956 (2008)
%

\bibitem{Coo14}
R.\ Cooke, M.\ Pettini, R.A.\ Jorgenson, M.T.\ Murphy, and C.C.\ Steidel,
Astrophys.\ J.\ {\bf 781}, 31 (2014)
%
\bibitem{Nollett:2011aa} 
  K.~M.~Nollett and G.~P.~Holder,
  %``An analysis of constraints on relativistic species from primordial nucleosynthesis and the cosmic microwave background,''
  arXiv:1112.2683 [astro-ph.CO]
  %%CITATION = ARXIV:1112.2683;%%

%\cite{DiValentino:2014cta}
\bibitem{DiValentino:2014cta} 
  E.~Di Valentino, C.~Gustavino, J.~Lesgourgues, G.~Mangano, A.~Melchiorri, G.~Miele and O.~Pisanti,
  %``Probing nuclear rates with Planck and BICEP2,''
  Phys.\ Rev.\ D {\bf 90}, 023543 (2014)
  %%CITATION = ARXIV:1404.7848;%%

\bibitem{LUNA02}
C.\ Casella {\it et al.}, (LUNA Collaboration), Nucl.\ Phys.\ A {\bf 706}, 
203 (2002)
%
\bibitem{Ma97}
L.\ Ma, H.\ Karwowski, C.\ Brune, Z.\ Ayer, T.\ Black, J.\ Blackmon, 
E.\ Ludwig, M.\ Viviani, A.\ Kievsky, and R.\ Schiavilla, 
Phys.\ Rev.\ {\bf 55}, 588 (1997)

%
\bibitem{Ade11}
E.G.\ Adelberger {\it et al.}, Rev.\ Mod.\ Phys.\ {\bf 83}, 195 (2011)
%
\bibitem{TrezziPV}
D.\ Trezzi for the LUNA Collaboration, private communication
%
\bibitem{Viv00}
M.\ Viviani, A.\ Kievsky, L.E.\ Marcucci, S.\ Rosati, and 
R.\ Schiavilla, Phys.\ Rev.\ C {\bf 61}, 064001 (2000)
%
\bibitem{Mar05}
L.E.\ Marcucci, M.\ Viviani, R.\ Schiavilla, A.\ Kievsky, and S.\ Rosati,
Phys.\ Rev.\ C {\bf 72}, 014001 (2005)
%
\bibitem{Wir95}
R.B.\ Wiringa, V.G.J.\ Stoks, and R.\ Schiavilla,
Phys.\ Rev.\ C {\bf 51}, 38 (1995)
%
\bibitem{Pud97}
B.S.\ Pudliner, V.R.\ Pandharipande, J.\ Carlson, S.C.\ Pieper, 
R.B.\ Wiringa,
Phys.\ Rev.\ C {\bf 56}, 1720 (1997)
%
\bibitem{Kie08}
A.\ Kievsky, S.\ Rosati, M.\ Viviani, L.E.\ Marcucci, L.\ Girlanda,
J.\ Phys.\ G {\bf 35}, 063101 (2008)
%
\bibitem{Gir10}
L.\ Girlanda, A.\ Kievsky, L.E.\ Marcucci, S.\ Pastore, R.\ Schiavilla, 
M.\ Viviani, 
Phys.\ Rev.\ Lett.\ {\bf 105}, 232502 (2010)
%
%\cite{Cyburt:2015mya}
\bibitem{Cyburt:2015mya} 
  R.H.~Cyburt, B.D.~Fields, K.A.~Olive and T.H.~Yeh,
  %``Big Bang Nucleosynthesis: 2015,''
  arXiv:1505.01076 [astro-ph.CO]
  %%CITATION = ARXIV:1505.01076;%%
%
\bibitem{Pas09}
S.\ Pastore, L.\ Girlanda, R.\ Schiavilla, M.\ Viviani, and R.B.\ Wiringa,
Phys.\ Rev. C {\bf 80}, 034004 (2009)

\bibitem{Son09}
Y.-H. Song, R. Lazauskas, and T.-S. Park, Phys. Rev. C {\bf 79}, 064002 (2009)
%
\bibitem{Gri62}
G.M.\ Griffiths, M.\ Lal, C.D.\ Scarfe,
Can.\ J.\ Phys.\ {\bf 41}, 724 (1962)
%
\bibitem{Sch96}
G.J.\ Schmid {\it et al.}, Phys.\ Rev.\ Lett.\ {\bf 76}, 3088 (1996)
%
\bibitem{Pia13}
M.\ Piarulli {\it et al.},
Phys.\ Rev.\ C {\bf 87}, 014006 (2013)
%
\bibitem{Mar15}
L.E.\ Marcucci {\it et al.}, 
 arXiv:1504.05063 [nucl-th]
%
\end{thebibliography}
\end{document}